\documentclass{iaus}
\ifx\pdftexversion\undefined
  \usepackage{epsfig,natbib}             %
\else
  \pdfoutput=1
  \usepackage{epsfig,natbib,hyperref}             %
  \hypersetup{pdfauthor={K. Petrovay},pdftitle={Solar and planetary dynamos}}
\fi

\renewcommand{\theequation}{\arabic{equation}}

\newcommand{\citen}[1]{\citeauthor{#1} \citeyear{#1}}

\newcommand{\areps}{{Ann.\ Rev.\ Earth Pl.\ Sci.\ }}		
		
\newcommand{\apj}{{ApJ\ }}			
\newcommand{\apjl}{{ApJ Lett.\ }}		
\newcommand{\aap}{{A\&A\ }}		
\newcommand{\nat}{{Nature\ }}

\newcommand{\Ek}{\mathrm{Ek}\,}
\newcommand{\Ra}{\mathrm{Ra}\,}
\renewcommand{\Pr}{\mathrm{Pr}\,}
\renewcommand{\Re}{\mathrm{Re}\,}
\newcommand{\Ro}{\mathrm{Ro}\,}
\newcommand{\Lo}{\mathrm{Lo}\,}
\newcommand{\Rab}{\mathrm{Ra}_{\mathrm{b}}}
\newcommand{\Rem}{\mathrm{Re}_{\mathrm{m}}}
\newcommand{\Ekm}{\mathrm{Ek}_{\mathrm{m}}}
\newcommand{\Remcrit}{\mathrm{Re}_{\mathrm{m,crit}}}
\newcommand{\LoD}{\mathrm{Lo}_{\mathrm{D}}}
\newcommand{\Prm}{\mathrm{Pr}_{\mathrm{m}}}
\newcommand{\Rol}{\mathrm{Ro}_{\mathrm{l}}}
\newcommand{\Lol}{\mathrm{Lo}_{\mathrm{l}}}
\newcommand{\rin}{r_{\mathrm{in}}}
\newcommand{\rout}{r_{\mathrm{out}}}
\newcommand{\varv}{v}

\title[Solar and Planetary Dynamos] 
{Solar and planetary dynamos:\\ comparison and recent developments}

\author[Petrovay]   
{K. Petrovay}

\affiliation{E\"otv\"os University, Department of Astronomy\\
H-1518 Budapest, Pf.~32., Hungary \\ 
email: {\tt K.Petrovay@astro.elte.hu}} 

\pubyear{2009}
\volume{257}  
\pagerange{1--11}
\jname{Universal Heliophysical Processes}
\editors{N. Gopalswamy et al.}

\begin{document}

\maketitle

\begin{abstract}
While obviously having a common root, solar and planetary dynamo theory have
taken increasingly divergent routes in the last two or three decades, and there
are probably few experts now who can claim to be equally versed in both.
Characteristically, even in the fine and comprehensive book ``The magnetic
Universe'' (\citen{Rudiger+Hollerbach}), the chapters on planets and on the Sun
were written by different authors. Separate reviews written on the two topics
include \cite{Petrovay:SOLSPA}, \cite{Charbonneau:livingreview},
\cite{Choudhuri:standard.review} on the solar dynamo and
\cite{Glatzmaier:AREPS}, \cite{Stevenson:pl.mgfields} on the planetary dynamo.
In the following I will try to make a systematic comparison between solar and
planetary dynamos, presenting analogies and differences, and highlighting some
interesting recent results.
\end{abstract}

\firstsection 

\section{Approaches to astrophysical dynamos}

\subsection{Dimensional analysis: mixing-length vs.\ magnetostrophic balance}
Faced with a problem like the dynamo, where the governing equations are well
known and the source of difficulties is their complexity, it is advisable to
start by order of magnitude estimates of the individual terms in the equations.
A clear and detailed account of this is given in \cite{Starchenko+Jones}.

In the case of the Sun, such considerations point to a balance between {\it
buoyancy} and {\it inertial forces} as the determinant of the resulting flow
pattern in the convective zone. Indeed, the order of magnitude equality of
these terms is one of the basic formul{\ae} of the mixing length theory of
astrophysical convection, as formulated in the 1950s ---so this balance is
customarily referred to as ``mixing length balance'' in the dynamo literature.

On the other hand, in rapidly rotating rotating systems the Coriolis term 
dominates over the inertial term. If the magnetic field is strong enough for
Lorentz forces to be comparable to the Coriolis force and the buoyancy, another
type of balance known as {\it magnetostrophic balance} (or MAC balance) sets in.
This kind of balance is now generally thought to prevail in the more
``mainstream'' planetary dynamos, such as those of Earth, Jupiter, Saturn, and
possibly Ganymede.

\subsection{Implicit models: Mean field theory}
As the smallest and largest structures present in the strongly turbulent
astrophysical dynamos are separated by many orders of magnitude, it is hopeless
and perhaps unnecessary to set the full explicit treatment of all scales of
motion as a goal. In this sense, all models of astrophysical dynamos are
necessarily ``mean field models'', not resolving scales smaller than a certain
level and representing the effect of those scales with some effective
diffusivities. Yet the term ``mean field theory'' is customarily reserved for
those models where even the largest scale turbulent motions, thought to be the
main contributors to dynamo action, remain unresolved. 

Mean field theory has remained the preferred theoretical tool of solar dynamo
studies. Even the effect of obviously non-mean-field effects can be included in
mean field models in the form of parametrized ad hoc terms. E.g. the emergence
of strong magnetic flux loops from the tacholine to the photosphere is now
thought by some to be a major contributor to the $\alpha$-effect (the so-called
Babcock-Leighton mechanism for $\alpha$). The motion of magnetized fluid being
highly independent of the rest of the plasma, this may appear to be a case
where the mean field description is bound to fail ---but the problem is
circumvented in mean field dynamo models e.g. by the introduction of a
non-local $\alpha$-term (\citen{Wang+:1.5D}) or by using different diffusivity
values for poloidal and toroidal fields (\citen{Chatterjee+:fullspheredyn}).

\subsection{Explicit models: Numerical simulations} In contrast to solar dynamo
theory, planetary dynamo studies have traditionally focused on models where the
large scale, rotationally influenced turbulent motions are explicitly resolved.
For a long time such studies were essentially kinematical, prescribing the large
scale flows by simple mathematical formul{\ae} that satisfy some more or less
well founded basic physical expectations, and solving only the induction
equation. Such studies can still significantly contribute to our understanding
of dynamos (see e.g. \citen{Gubbins:kinematic.geodynamo} for a recent overview).
Yet, from the mid-1990s onwards the rapid increase in computer power has made it
possible to develop explicit dynamical models (aka numerical simulations) of the
geodynamo, and this has become the main trend in planetary dynamo research.

\section{Observational constraints}

We are separated from the Earth's outer core by 3000 km of intransparent rock,
while the top of Sun's convective zone is directly observable across 1 AU of
near-empty space. Although magnetic measurements and seismology in principle
allow indirect inferences on conditions in the outer core, these are both
limited to relatively large-scale (spherical harmonic degree $l<13$) magnetic
structures, while empirical information on flows in the core is almost
completely lacking.

In sharp contrast to this, the brightness of the Sun allows high S/N detection
of spectral line profiles, permitting a precise determination of Doppler and
Zeeman shifts. This not only provides a wealth of high-resolution observations
of flows and magnetic fields at the top of the convective zone: the sensitive
detection of waves and oscillations in the solar photosphere also allows a 
detailed reconstruction of flows and magnetic field patterns in layers lying
below the surface. 

The different amount of empirical constraints are certainly a major factor in
determining the different approaches taken by solar and planetary dynamo
studies. The main shortcoming of mean field models impeding progress is the vast
number of possibilities available for the choice of parameters and their
profiles, which renders the formulation of mean field dynamo models for the
planets a rather idle enterprise. The good empirical constraints in the solar
case provide an indispensable support by narrowing down the range of admissible
models. It is no coincidence that it was the helioseismic determination of the
internal rotation profile around 1990 which led to a resurgence in solar mean
field dynamo theory.

At the same time, no MHD numerical simulation of the solar convective zone has
been able to recover the observed butterfly diagram, and even reproducing the
observed differential rotation profile is not trivial (cf. Sect.~6 below). 
Geodynamo simulations correctly reproduce nearly all the known spatiotemporal
variance in the Earth's magnetic field, and recently there are indications that
they could provide a successful general scheme to understand the variety of
dynamos seen in the Solar System (see Sect.~8).

\section{State of matter, stratification}

One obvious difference between planetary and stellar dynamos is that while the
conducting matter in stars is ionised gas, in planetary dynamos it's conducting
liquids. The consequences of this are twofold. On the one hand, in planets
incompressibility is often a good approximation, whereas in the Sun
there many scale heights between the top and bottom of the SCZ
and the scale height itself varies by several orders of magnitude. This extreme stratification or
``stacking'' of structures of vastly different scale is one major obstacle in
the way of realistic global simulations of the SCZ.

Another consequence of the different state of matter in stellar and planetary
dynamos is that molecular transport coefficients or diffusivities (such as
viscosity, resistivity or heat diffusivity) are less accurately and reliably
known in planetary interiors, given the more complex material structure.
Together with the unsatisfactorily constrained thermal state, this means that
the exact position, boundaries and even nature of the dynamo shell is in doubt
in some planets. In the water giants (Uranus and Neptune) the resistivity of
their electrolytic mantles is very uncertain, affecting the extent and depth of
the dynamo layer. In Mercury, the unknown thermal state of the planet and the
unknown amount of light constituents make the position and thickness of the
liquid outer core rather uncertain. (Evidence for a liquid outer core was
recently reported by \citen{Margot+:Mercury.moltencore}.)

\section{Energetics and importance of chemistry}

The energy source of the convective motions giving rise to the dynamo
introduces further divisions into astrophysical dynamos. Jupiter, Saturn and
the Sun are luminous enough to drive vigorous convection in their interiors by
thermal effects alone. For terrestrial planets the available remanent heat
may be only barely sufficient or even insufficient to maintain core convection.
In these objects, chemical or compositional driving may be an important
contribution to maintaining the convective state of their cores. (See e.g.
\citen{Stevenson:pl.mgfields} for numerical estimates.) In the case of the
Earth, the most likely candidate is the piling up of light constituents like
sulphur at the bottom of the outer core, as iron is freezing out onto the inner
core and the light elements are locked out of the solid phase. (Arguments for
an analoguous mechanism suggested for Saturn, the ``helium rain'' now seem to
have been weakened by the new results of \citen{Stixrude+:liquidHe}.)

\section{Boundaries and adjacent conducting flows}

One important new contribution of numerical simulations was the realization of
the crucial role that the choice of boundary conditions play in determining the
solutions. Indications for this effect had already been found in kinematic
dynamo calculations: e.g. the assumed conductivity of the solid inner core in a
geodynamo model affects strongly its capability to maintain a dipole dominated
field and the frequency of reversals (\citen{Hollerbach+:innercore}). But the
most important effect is due to the thermal boundary conditions. Convection is
driven by heat input from below and heat loss from the top of the layer. In
terrestrial planets the latter occurs by mantle convection, the efficiency of
which is then a fundamental determinant of the behaviour of the dynamo. Indeed,
as the timescale of mantle flows is very long compared to those in the dynamo,
in addition to the overall mantle convection even the instantaneous convection
pattern realized now can have a profound effect on the flow structure in the
geodynamo (\citen{Glatzmaier:AREPS}). The most widely mentioned possibility why
Venus does not seem to support a dynamo is that, lacking plate tectonics,
Venus' mantle convection is less efficient, so its heat flux can be transported
conductively throughout the core, and no convective instability arises. The
resulting slower cooling of the planet may have the additional result that no
inner core has solidified yet, depriving Venus even of the alternative energy
source for convection that an inverse molecular weight gradient could provide
(\citen{Stevenson:pl.mgfields}).

The importance of appropriate boundary conditions is now also recognized in
simulations of the solar convective zone. Despite repeated attempts with ever
more powerful computers, the hydrodynamical simulations of the new millennium
had until very recently not been able to reproduce the observed solar internal
rotation profile, the resulting isorotation surfaces being cylindrical rather
than conical (e.g.\ \citen{Brun+:cylindrical}). The most recent
state-of-the-art simulation (\citen{Miesch+:conical}), however, brought a
breakthrough in this respect. The breakthrough is not due to the higher
computing power available, but to the introduction of a small ($\sim0.1\,$\%)
pole-equator temperature difference at the bottom boundary. While at first
sight ad hoc, the introduction of this difference had the beneficial effect of
leading to a realistic differential rotation and a realistic meridional
circulation at the same time, raising hopes that this may indeed be the right
way out of the dilemma. This is so despite the fact that this solution of the
problem may strike one as ``{\ae}sthetically'' less attractive, as the conical
isorotation surfaces are attributed to a chance cancellation of two opposing
effects, rather than to some deeper physical motive.

A further important factor that may fundamentally affect the behaviour of a
dynamo is the presence or absence of conducting stable fluid layers in its
vicinity. Such a layer, the solar tachocline, is certainly present below the 
solar convective zone, and is currently thought to play a major role in the
global solar dynamo (\citen{FDE+Py:fast1}; \citen{FDE:fast2}). The strong
stratification of the solar convective zone discussed above, has long been
expected to give rise to pronounced transport effects or ``pumping'' mechanisms
(see \citen{Petrovay:NATO} for a comprehensive treatment of these effects).
These pumping processes can efficiently remove most of the large-scale magnetic
flux from the turbulent region, provided there is an adjacent region with high
but finite conductivity that can receive and store this flux. Following earlier
simpler numerical experiments, recent MHD numerical simulations have indeed
shown the pumping of large scale magnetic flux from the convective zone into
the tachocline below, where it forms strong coherent toroidal fields
(\citen{Browning+:dynsimu.pumping}).

Similar conductive layers adjacent to the turbulent dynamo region may be
present in some planetary interiors. Even in the case Earth it has been
hypothesized that, as a result of convection, lighter elements ultimately pile
up below the the CMB, resulting in a stably stratified sublayer within the
outer core (\citen{Whaler}). The effect of such layers on the dynamo has not
been studied extensively, but available results indicate that their impact on
the observed field may be very important (\citen{Schubert+:overlying.flows}).

\section{Regime of operation}

\subsection{Dimensional and nondimensional parameters} The number of parameters
uniquely determining a dynamo is quite limited. The geometry is a spherical
shell between radii $\rin$ and $\rout$, with thickness $d=\rout-\rin$ and
relative thickness parameter $x=d/\rout$. The shell, rotating with angular
velocity $\Omega$, is filled with material of density $\rho$, characterized by
momentum, heat and magnetic diffusivities $\nu$, $\kappa$ and $\eta$,
respectively. Finally, convection in the shell is driven by the buoyancy flux
$F$ fed in at the bottom of the shell
(\citen{Olson+Christensen:pldynamo.scaling}). An internal heating $\epsilon$
may be added to the list for cases where volumetric heat loss from secular
cooling or radioactive decay is significant.

The number of relevant parameters can be reduced further, realizing that the
role of some of the variables is just to set characteristic scales. The length
scale is clearly set by $d$ and the mass cale by $\rho$. For the timescale it
has been traditional for certain theoretical considerations to use the
resistive time $d^2/\eta$  in dynamo theory. However, it was recently pointed
out by \cite{Christensen+Aubert:pldynamo.scaling} that for rapid rotators
$1/\Omega$ offers a more relevant scaling. Nondimensionalizing all parameters
by these scales (and ignoring $\epsilon$) we are left with only 4
nondimensional parameters. Following \cite{Olson+Christensen:pldynamo.scaling}
the effective  buoyant Rayleigh number can be defined as $\Rab=
F/(1-x)d^2\Omega^3$. The nondimensional measures of the diffusivities are the
Ekman numbers
\begin{equation} \Ek=\nu/\Omega d^2 \qquad \mathrm{Ek}_\kappa=\kappa/\Omega d^2 \qquad
\Ekm=\eta/\Omega d^2 \end{equation} 
Instead of the three Ekman numbers, one Ekman number and two Prandtl numbers
(i.e. diffusivity ratios) are more commonly used:
\begin{equation} \Pr=\nu/\kappa \qquad \Prm=\nu/\eta \end{equation} 

The specified parameters then, in principle, determine the solution, i.e. the
resulting turbulent flow field and magnetic field. These can be characterized by
their respective amplitudes $v$ and $B$, as well as their typical length scale
$l$ ---say correlation length, integral scale or similar. (For simplicity a
single length scale is assumed for both variables, ignoring anisotropy).  In
fact, these fiducial scales of the solution may even be estimated on dimensional
grounds without actually solving the dynamo equations
(\citen{Starchenko+Jones}). From $l$ and $v$, turbulent diffusion coefficients
can be estimated using the usual {\it Austausch} recipe $\sim lv$. For the Sun,
Earth and gas giants such estimates, summarized in Table 1, are in rough
agreement with both observations and simulation results.

As nondimensional measures of $v$ and $B$ we introduce the
Rossby and Lorentz numbers:
\begin{equation} 
  \Ro=\varv/\Omega d \qquad  \Lo=\varv_A/\Omega d 
\end{equation} 
where $v_A=B/(\rho\mu)^{1/2}$ is the Alfv\'en speed. It arguably makes more
sense to use $l$ instead of $d$ in the above definitions, which leads us to
``local'' versions of these numbers:
\begin{equation} \Rol=\varv/\Omega l \qquad  \Lol=\varv_A/\Omega l 
\end{equation} 

Were we to adopt the diffusive time scales $d^2/\nu$ and $d^2/\eta$ instead of
$1/\Omega$, the nondimensional measures of $v$ and $B$ would be the kinetic and
magnetic Reynolds numbers:
\begin{equation} 
\Re=d\varv/\nu  \qquad  \Rem= d\varv/\eta
\end{equation} 
(Again, using a ``local'' length parameter $l$ instead of $d$ may be more
relevant.)

\begin{table}[ht]
\begin{center}
\caption{Estimated  values of the flow velocity, length
scale and turbulent diffusivity}
\begin{tabular}{lrrr}
\hline
Dynamo & $\varv$ [m/s] & $l$ [km] & $D_t\sim l\varv$ [m$^2$/s] \\
\hline
   Sun (deep SCZ): & $20$ & $10^5$ & $10^9$ \\
   Earth: & $3\cdot 10^{-4}$ & $100$ & 20 \\
   Jupiter & $10^{-3}$--$10^{-2}$ & $10^3$ & $10^3$\\
\hline
\end{tabular}
\end{center}
\end{table}

Let us now consider what the characteristic values of these nondimensional
parameters are for astrophysical dynamos and for numerical simulations.

\subsection{Ekman numbers (i.e. diffusivities)}
Ekman numbers are extremely small in astrophysical dynamos, far below the range
$10^{-6}$--$10^{-3}$ accessible to current numerical simulations. The main
obstacle to the further reduction of $\Ek$ is that by choosing a realistic
Rayleigh number, we fix the input of kinetic and magnetic energy into the
system. In a stationary state energy must be dissipated at the same rate as it
is fed in. The viscous dissipation rate is $\sim\nu/\lambda^2$ where $\lambda$
is the smallest resolved scale ---so $\nu$ cannot be reduced without also
increasing the spatial resolution. The situation is similar for resistive
(Ohmic) dissipation. 

\subsection{Reynolds numbers}
It is interesting to note that the magnetic Reynolds number $\Rem$ is the only
parameter involving a diffusivity whose actual value can be used in current
numerical simulations for some planets (including the Earth). Rossby numbers,
in contrast, are invariably intractably small in planetary dynamos. The
situation in the Sun is the reverse: Rossby numbers are moderate (only slightly
below unity in the deep convective zone and actually quite high in shallow
layers). Reynolds numbers, however, are all exteremely high in the solar
plasma, even in the relatively cool photosphere.

The significance of $\Rem$ consists in the existence of a critical value
$\Remcrit$ below which no dynamo action is possible. Planetary dynamo
simulations have led to the surprising result that $\Remcrit$ has a universal
value of about 40, independently of the other parameters, specifically of the
Prandtl numbers (\citen{Christensen+Aubert:pldynamo.scaling}). This is a
surprising result as the expectation in turbulence theory, confirmed in
numerical simulations of small-scale dynamos, has been that $\Remcrit$ should
be a sensitive function of $\Prm$ (\citen{Boldyrev+Cattaneo:Rmcrit}). The
solution of this apparent contradiction is not known.

Elementary estimates show that the convective velocity resulting even for very
slightly supercritical Rayleigh numbers results in magnetic Reynolds numbers
well above $\Remcrit$. This implies that the conditions for convection and for
dynamo action may be virtually identical in astrophysical fluid bodies with a
high conductivity (\citen{Stevenson:pl.mgfields}). We have already seen that
the apparent lack of dynamo action in Venus is attributed to the lack of
convection in its core. Planets with very poorly conductive fluid layers,
however, may obviously be convecting without supporting a dynamo. A case in
point may be the water giants where the dynamo sustaining layer may potentially
only extend to a thin sublayer of the convecting water mantle and/or be only
slightly supercritical, unable to generate fields strong enough to reach
magnetostrophic equilibrium.

\subsection{Rossby number (i.e. velocity amplitude)}
An important issue is how the Rossby and Lorentz numbers (i.e. nondimensional
velocity and magnetic field amplitudes) scale with the input parameters of the
dynamo problem. The significance of this is twofold. Firstly, as current
numerical simulations cannot directly access the parameter range relevant for
astrophysical dynamos, such scaling laws can be used to extrapolate their
results into the physically interesting domain. Secondly, it is of interest to
compare the scaling laws derived from simulation results to those predicted by
physical considerations based on the concept of magnetostrophic equilibrium.

For the Rossby number, the scaling law extracted from simulations vs. the law
theoretically expected for MAG balance are:
\begin{equation}
  \Ro=0.85\, {\mathrm{Ra}_{\mathrm{b}}^{0.4}} \qquad 
  \mathrm{vs.}  \qquad \Ro\sim{\mathrm{Ra}_{\mathrm{b}}^{1/2}} 
  \label{eq:Rossby}
\end{equation}

Geometrical factors invoked to explain the slight discrepancy in the values of
the exponents do not seem to be capable of explaining it
(\citen{Christensen+Aubert:pldynamo.scaling}). However, given the above
mentioned doubt that the high values of the diffusivities employed may cast on
all simulation results, one may also take the point of view that these two laws
are actually in fairly good agreement, and no further explanation is needed.
(Note that in this case the theoretical value 0.5 must be deemed the more
reliable one.)

\subsection{Lorentz number (i.e. magnetic field strength)}

The scaling of the Lorentz number extracted from dipole-dominated dynamo
simulations with no internal heating
(\citen{Olson+Christensen:pldynamo.scaling}) viz. the MAC scaling following
from the assumption of unit Els\"asser number (\citen{Starchenko+Jones}) are
\begin{equation} 
  \LoD\propto {\mathrm{Ra}_{\mathrm{b}}^{1/3}} \qquad
  \Lo\propto {\mathrm{\Rol}}^{1/2} 
  \label{eq:Loscaling}
\end{equation} 
Assuming that the local Rossby number scales similarly to (\ref{eq:Rossby}),
the latter relation may also be turned into a scaling with $\Ra$, but its
exponent ($0.2$--$0.25$) is even more discrepant from the value yielded by the
simulations than in the previous case.

\section{Predictive value}

In solar dynamo theory there is a widely held belief that predictions for at
least the next activity cycle should be possible. Dozens of methods have been
proposed for this. The relatively most successful ones are apparently those
based on some measure of solar activity or magnetism at the onset of the new
cycle.Recently \cite{Cameron+Schussler} convincingly argued that what stands in
the background of all such  methods is just the well known Waldmeier effect,
relating a cycle's amplitude to the rate of rise of activity towards the
maximum. As solar cycles are known to overlap by $\sim1$--$2$ years, the faster
rise preceding a stronger cycle will result in an earlier epoch for the minimum
---so it is natural to expect that at this time, polar magnetic fields and
activity indices have not yet decayed to such low values as they reach during
minima preceding weak cycles. Indeed, \cite{Cameron+Schussler} demonstrate that
a very simple predictor, the value of the sunspot number three years before the
minimum performs embarrassingly well (correlation coefficient 0.95) when it
comes to predicting the amplitude of the next maximum, but this good
performance is fully explained by the effect described above.

This is not to say that it would be pointless to rely on dynamo models to find
a physical basis for activity predictions. Firstly, the Waldmeier effect
itself is certainly due to some, as yet unclear, aspect of the dynamo. Second,
there are grounds to assume that, at least in certain types of dynamos, there
really {\it is} a physical relationship between high-latitude magnetic fields
during the minimum and the amplitude of the next maximum
(\citen{Choudhuri+:prediction}). Either way, the persistently low activity
during the present solar minimum seems to be a strong indication that cycle 24
will be a rather weak one, in contrast to  widely publicized claims based on  a
certain class of solar dynamo models (\citen{Dikpati+:prediction}).

Owing to the much longer timescales of planetary dynamos, the possibility for
testable temporal predictions is limited. Planetary dynamo models can partly
make up for this by the availability of several instances of known planetary
dynamos. These inlcude 7 active dynamos (Mercury, Earth, Ganymede and the four
giant planets) and one extinct dynamo (Mars), with the remaining planets  also
providing some important constraints precisely by apparently {\it not}
supporting a dynamo. A testbed of a certain type of modelling approach, then,
may be whether it is capable to provide a unified explanatory scheme for all
these systems. Recent results suggest that planetary dynamo simulations may
indeed provide such an explanatory scheme.

\begin{figure}
\begin{center}
\epsfig{figure=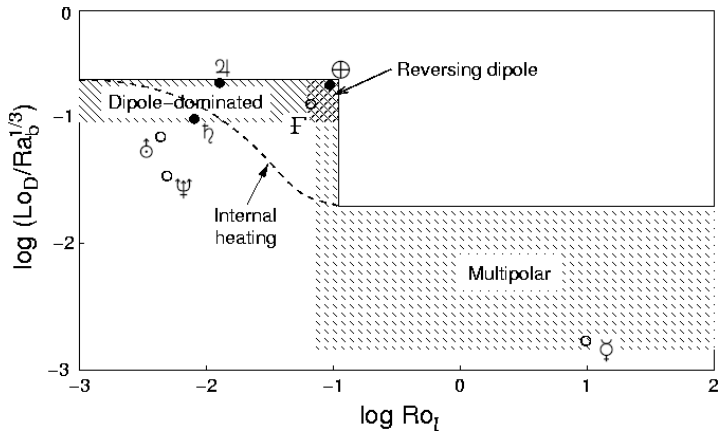,width=0.7\textwidth}
\end{center}
\caption{The suggested unified classification scheme of planetary dynamos based
on the scalings of \cite{Olson+Christensen:pldynamo.scaling}}
\end{figure}

Collecting and homogenizing the results of hundreds of geodynamo simulations, 
\cite{Olson+Christensen:pldynamo.scaling} find an interesting scaling (or
rather, ``non-scaling'') behaviour of the amplitude of the dipolar component of
the resulting magnetic field. As mentioned above, considering only
dipole-dominated dynamos without internal heating the scaling relationship
(\ref{eq:Loscaling}) results for the amplitude of the dipole component. This
scaling, however, is not valid for all dynamo models. This is borne out in
Figure~1 where the ordinate essentially shows the ratio of the two sides of
equation (\ref{eq:Loscaling}). The dipole-dominated cases in this plot will
clearly lie along a horizontal line drawn at the ordinate value 0.5,
corresponding to the coefficient in equation (\ref{eq:Loscaling}). This is
indeed the case for dynamos where the Rossby number is low, i.e. where the
driving is relatively weak. However, for Rossby numbers above a critical value
of order 0.1, the amplitude of the dipolar component suddenly drops and the
solution becomes multipolar. In the case of internally heated models the
situation is similar, but the transition between the two regimes is more
gradual. 

The critical role of the Rossby number in this respect indicates that the
underlying cause of the eventual collapse of the dynamo field is the increasing
importance of inertial forces in the equation of motion (measured by the Rossby
number). This leads to a breakup of the relatively regular columnar structures
dominating rapidly rotating convection ---the detailed mechanism of this will
be discussed further in the next section.

A further interesting finding is that dynamos lying near the top of the Rossby
number range of dipolar solutions are generally dipole-dominated, but
occasionally undergo excursions and reversals. In a turbulent system this type
of behaviour is rather plausible, given that parameters like $\Ro$ are expected
to fluctuate, and such fluctuations may occasionally take the system into
multipolar regime, causing the dipolar field to collapse, and then reform once
fluctuations have taken the system back to the dipolar regime.

\cite{Olson+Christensen:pldynamo.scaling} also attempt to place individual
planetary dynamos on the phase plane of Fig.~1, based on known empirical
constraints and theoretical considerations. The resulting distribution seems to
provide an impressively comprehensive classification system for planetary
dynamos. Gas giants lie safely in the dipole-dominated regime, suggesting that
their dynamos are in a magnetostrophic state maintaining a nonreversing dipolar
field. Earth is found near the limit of the dipole-dominated regime, just where
a dipolar dynamo known to be subject to occasional reversals is expected to be.
A somewhat surprising conclusion of this scheme is that the planetary dynamo
most similar to Earth's in its behaviour may be Ganymede's, also expected to
maintain a reversing dipole. The fact the dipolar magnetic field amplitude of
the water giants is significantly less than  relations (\ref{eq:Loscaling})
predict, either be due to a significant amount of internal heating or to the
low resistivity in their interiors, which prevents them from reaching
magnetostrophic equilibrium. Finally, Fig.~1 would seems to suggest that,
despite contrary opinions, Mercury should be expected to sustain a multipolar
magnetic field structure.

Some caveats regarding Fig.~1 are in order. All the simulations upon
which the figure is based were run with a shell thickness parameter $x=0.65$,
appropriate for the geodynamo. For planets with significantly different shell
geometries, especially for those with thin shells (Mercury?; the water
giants?), results may prove to be  significantly different. 

Note also that the parameter on the abscissa of Fig.~1 is actually a rather
particular kind of Rossby number, involving one particular local turbulence
timescale, defined in a way similar to the the Taylor microscale. As in current
numerical simulations the largest and smallest scales are normally
separated by no more than one order of magnitude, this fine distinction between
different turbulent scales is not important. But when it comes to
application to actual planets, the different turbulent timescales differ by
many orders of magnitude, so choosing the right scale is critical.
\cite{Christensen+Aubert:pldynamo.scaling} provide good arguments for the
choice of the particular form of $\Rol$ used in Fig.~1, but it is still
somewhat disconcerting that combining large-scale turbulent velocities with a
much smaller length scale, the Rossby number loses its widely used physical
interpretation as the ratio of rotational and turbulent turnover timescales. In
addition, our limited knowledge on the spectrum of magnetostrophic turbulence
(\citen{Zhang+:AREPS}, \citen{Nataf+Gagniere:VKS}) makes it hard to derive a
reliable value for the Taylor microscale in planetary dynamos: the Kolmogorov
spectrum is just a (not too educated) guess.

\section{Characteristic flow and field patterns}

It is well known that solar activity phenomena appear collectively, in the form
{\it active regions}. These regions are the solar atmospheric manifestations of
large azimuthally oriented magnetic flux loops emerging through the convective
zone. Indeed, the development of extremely successful detailed models of this
emergence process, first in the thin fluxtube approximation, then in 3D
numerical simulations, was probably the most spectacular success story of solar
dynamo theory in the last few decades. 

In contrast to our familiarity with the basic magnetic structures determining
solar activity, nothing about analoguous structures in planetary dynamos had
been known until very recently. This situation has spectacularly changed with
the development of new visualization and geometrical analysis techniques, which
showed that, just like in the Sun, the magnetic field in planetary dynamos is
highly intermittent (25\,\% of the magnetic energy residing in just 1.6\,\% of
the volume), and led to the recognition of a variety of field and flow
structures  (\citen{Aubert+:geodyn.flowstruct}). The two main classes of such
structures are {\it magnetic cyclones/anticyclones} and {\it magnetic
upwellings}.

Magnetic cyclones and anticyclones are the MHD equivalent of
Busse's columnar structures, known to dominate rapidly rotating 
hydrodynamic convection. The simulations indicate that magnetic anticyclones, in
particular, play a key role in generating and maintaining a dipole dominated
magnetic field in the upper part of the dynamo layer and above. Near the bottom
of the shell the field is invariably multipolar, but the thermal wind-driven
upflow in the axis of a magnetic anticyclone amplifies the upward convected
fields by stretching, so that at the top of the shell a predominantly dipolar
field results.

Magnetic upwellings are buoyancy-driven upflows rising nearly radially through
the shell owing to their high velocity. They come in two varieties: polar
upwellings, limited to the interior of the tangent cylinder of the inner core,
emerge more or less in parallel with the cyclonic/anticyclonic structures,
while equatorial upwellings cut through those structures, disrupting their
integrity. As a result, equatorial upwellings (distant cousines of the emerging
magnetic flux loops in the Sun) are capable of interfering with the maintenance
of an organized dipolar field by the magnetic anticyclones. As the buoyant
driving, and in consequence the turbulent velocity (i.e. the Rossby number) is
increased, the number and vigour of upwellings increases, until they can
occasionally disrupt the dipolar magnetic field maintained by the magnetic
anticyclones. Following such interruptions the dipole may be reformed with a
polarity parallel or opposite to its previous polarity: such events correspond
to magnetic excursions and reversals, respectively. Finally, with the further
increase of $\Ro$ the equatorial magnetic upflows permanently discapacitate the
anticyclones, and the dominant dipolar field structure collapses. This is the
mechanism in the background of the characteristic $\Ro$-dependence of dynamo
configurations shown in Fig.~1 and discussed in the previous section. Thus,
dynamo studies now seem to have elucidated, at least on a qualitative level,
the basic mechanisms underlying the most characteristic phenomena of both solar
and planetary dynamos.


\acknowledgement
Support by the Hungarian Science Research Fund (OTKA) under grant no.\ K67746 
and by the EC SOLAIRE Network
(MTRN-CT-2006-035484) is gratefully acknowledged.




\end{document}